\documentclass{aa}
\usepackage[varg]{txfonts}
\usepackage{graphicx}
\begin{document}

\title{ANALYSIS OF A CUSTOM SUPPORT VECTOR MACHINE FOR PHOTOMETRIC REDSHIFT ESTIMATION AND THE INCLUSION OF GALAXY SHAPE INFORMATION}

\author{E. Jones\thanks{evan.jones@richmond.edu}, \and J. Singal}

\institute{Physics Department, University of Richmond\\28 Westhampton Way, University of Richmond, VA 23173}

\abstract {} {
 We present a custom support vector machine classification package for photometric redshift estimation, including comparisons with other methods. We also explore the efficacy of including galaxy shape information in redshift estimation. Support vector machines, a type of machine learning, utilize optimization theory and supervised learning algorithms to construct predictive models based on the information content of data in a way that can treat different input features symmetrically, which can be a useful estimator of the information contained in additional features beyond photometry, such as those describing the morphology of galaxies.
}
{The custom support vector machine package we have developed is designated SPIDERz and made available to the community.  As test data for evaluating performance and comparison with other methods, we apply SPIDERz to four distinct data sets:  1) the publicly available portion of the PHAT-1 catalog based on the GOODS-N field with spectroscopic redshifts in the range $z < 3.6$, 2) 14365 galaxies from the COSMOS bright survey with photometric band magnitudes, morphology, and spectroscopic redshifts inside $z < 1.4$, 3) 3048 galaxies from the overlap of COSMOS photometry and morphology with 3D-HST spectroscopy extending to $z < 3.9$, and 4) 2612 galaxies with five-band photometric magnitudes and morphology from the All-wavelength Extended Groth Strip International Survey and $z < 1.57$.  }
 {We find that SPIDER-z achieves results competitive with other empirical packages on the PHAT-1 data, and performs quite well in estimating redshifts with the COSMOS and AEGIS data, including in the cases of a large redshift range ($0 < z < 3.9$).  We also determine from analyses with both the COSMOS and AEGIS data that the inclusion of morphological information does not have a statistically significant benefit for photometric redshift estimation with the techniques employed here.  } {}

\titlerunning{SVM photo-z}
\authorrunning{Jones \& Singal}

\maketitle

\section{Introduction} \label{intro}

An important challenge for the current and coming era of large multi-band extragalactic surveys is obtaining sufficiently accurate photometric redshift estimates and understanding the error properties of these estimates (see e.g. \citet{Huterer06} for a review).  Unlike time consuming spectroscopic redshift determination, photometric redshift estimation (photo-z) is subject to significant systematic errors and confusion because the spectral information of a galaxy is limited to the magnitude or flux in a number of wavelength bands.  When photo-zs are used, science goals such as using weak lensing for cosmology are strongly affected by the number of outliers --- those objects whose estimated photo-zs are far from the actual redshifts \citep[e.g.][]{Hearin10}.  In general, data sets with bands extending into those observed by infrared telescopes (e.g. J, H, and K bands) have more accurate photo-z estimation and fewer outliers.  However, most upcoming large surveys, such as the Large Synoptic Survey Telescope \citep[LSST,][]{LSSTover}, will have optical and near-infrared data only.  Reducing the number of, and potentially having a method of identifying, the potential outliers in photo-z estimation is an important goal for these projects.   

Photo-z estimation techniques have traditionally been divided into two main classifications.  So-called ``Template fitting'' methods, such as the {\it Lephare} package as described in \citet{lephare} and \citet{Arnouts99}, and {\it Bayesian Photometric Redshift (BPZ)} as described in \citet{bpz}, involve correlating the observed band photometry with model galaxy spectra and redshift, and possibly other model properties.  So-called ``Empirical'' or ``Training set'' methods, such as artificial neural networks \citep[e.g. {\it ANNz},][]{annz}, boosted decision trees \citep[e.g. {\it ArborZ},][]{Gerdes10}, regression trees / random forests \citep[e.g.][]{Carliles10,KB13}, support vector machines \citep[e.g.][]{W04}, polynomial mapping \citep[e.g.][]{B05,LY08}, and others develop a mapping from input parameters to redshift with a training set of data in which the actual spectroscopic redshifts are known, then apply the mappings to data for which the redshifts are to be estimated.  Both have their drawbacks --- template fitting methods require assumptions about intrinsic galaxy spectra or their redshift evolution, and empirical methods require the training set to be ``complete'' in the sense that it is representative of the target evaluation population in bulk in all characteristics.  In a previous work \citep{NNP} we reported on a custom artificial neural network algorithm for photometric redshift determination.

Because of the larger frequency of mergers at higher redshifts and the general evolutionary trend from spiral to elliptical shapes among galaxies, it is a reasonable hypothesis that galaxy morphology and redshift are correlated in such a way that the addition of morphological information could improve photo-z estimation.  The inclusion of morphological parameters in photo-z estimation has been studied using Sloan Digital Sky Survey (SDSS) data by \citet{Tag03} with an artificial neural network determination, and by \citet{VC2006} and \citet{WS2006} with other methods.  \citet{Tag03} find possible modest improvement with the inclusion of shape information, although they restrict their analysis to quite low redshift ($z \leq$ 0.7) galaxies. \citet{WS2006} consider several empirical methods and show marginal improvement for some methods with the addition of morphological information.  \citet{VC2006} claim an improvement of between 1 and 3 percent in the RMS error in photo-z determination, however it is not noted whether this result is significant and the method of photo-z estimation is not discussed.  \citet{NNP} considered one of the galaxy test data sets used here which includes a more thorough sample of higher redshift galaxies extending to $z \sim 1.57$ and found that including galaxy shape information did not result in a statistically significant benefit in the context of a neural network method.  

In this work, we evaluate the performance of a custom support vector machine (SVM) package for photo-z determination with comparisons to other photo-z algorithms, and also explore the efficacy of integrating parameters describing the morphological information of galaxies.  The SVM package used in this analysis is developed for the IDL environment by one of the authors (EJ), based largely on algorithmic procedures outlined in \citet{LIBSVM}, and has been named SPIDERz (SuPport vector classification for IDEntifying Redshifts).\footnote{available from http://spiderz.sourceforge.net with usage documentation provided there.}   It can include additional parameters beyond photometry, such as morphological information, on an equal footing.

Here we follow convention \citep[e.g.][]{Hildebrandt10} and define ``outliers'' as those galaxies where
\begin{eqnarray}
Outliers: {{\vert z_{phot}-z_{spec} \vert} \over {1+z_{spec}}} > .15, 
\label{erroreq}
\end{eqnarray}
where $z_{phot}$ and $z_{spec}$ are the estimated photo-z and actual (spectroscopically determined) redshift of the object.  The RMS photo-z error in a realization is given by a standard definition 
\begin{eqnarray}
\sigma_{\Delta z/(1+z)} \equiv \sqrt { {{1} \over {n_{gals}}}  \Sigma_{gals} \left( {{ z_{phot}-z_{spec} } \over {1+z_{spec}}} \right) ^2 }, 
\label{RMSeq}
\end{eqnarray}
where $n_{gals}$ is the number of galaxies in the evaluation set and $\Sigma_{gals}$ represents a sum over those galaxies. 
For comparison with other works, we also calculate in certain determinations the RMS error without the inclusion of outlier galaxies, referring to this quantity as the ``reduced'' RMS or R-RMS.

The paper is organized as follows.  In \S \ref{method} we discuss the SVM model implemented in SPIDERz.  In the following sections we perform analyses with four separate test data sets and make comparisons to other photo-z methods when possible.  In \S \ref{PHAT} we discuss the results of testing SPIDERz on the publicly available portion of the PHoto-z Accuracy Testing real galaxy data catalog (PHAT-1) which spans the redshift range $z < 3.6$ and provides a useful comparison with other photo-z estimation methods.  In \S \ref{COSMOS} we discuss the results of testing SPIDERz on the Cosmic Evolution Survey (COSMOS) bright catalog with 14365 galaxies spanning $z < 1.4$ with ten-band photometry, spectroscopic redshifts, and seven morphological measurements, highlighting results obtained with differing numbers of bands and the inclusion of morphological information.  We perform a similar analysis in  \S \ref{COS-HST} with a catalog of 3048 galaxies spanning a wider redshift range ($z < 3.9$), which were obtained by combining COSMOS photometry and morphology with 3D-HST spectroscopy. In \S \ref{AEGIS} we discuss the results of testing SPIDERz on a catalog of 2612 galaxies with spectroscopic redshifts, five-band photometry, and morphological information from the All-wavelength Extended Groth Strip International Survey (AEGIS) in the range $z < 1.57$, and make a comparison to a neural network determination on this data.  We present a summary in \S \ref{disc}. 

\section{Support vector machine photo-z estimation method} \label{method}

\subsection{Motivation for use}

Support vector machines are a popular statistical machine learning tool that have been successfully applied for a variety of applications \citep[e.g.][]{Hearst98}. Within astronomy and astrophysics, SVMs have been used in recent works for automated image separation \citep[e.g.][]{Beau11}, galaxy morphological classification \citep[e.g.][]{H-C08} and classification of objects into stellar, galactic, or active galaxy categories using morphology, colors, or spectra \citep{Marton16,Malek13,Hassan13,Solarz13,Klement11,Peng02}.  

In the case of photo-z analysis, the use of SVMs has been reported by \citet{W04} and \citet{Wang13}.  Both restricted consideration to lower redshift sources than the present analysis.  In the case of the later, the sources are of very low redshift (all $z < 0.5$ and most $z < 0.3$).  In the case of the former, analyses were restricted to $z < 1$ sources except in the case of a simulated set.

In SVM photo-z determinations, the output of the training algorithm is a mapping from band magnitudes and, in principle, other information such as shape parameters, to redshift. The SVM learning algorithms employed in SPIDERz treat the band magnitudes and other information equally, making it a useful tool for exploring whether additional parameters beyond the band magnitudes, such as morphology in this case, provide additional useful information. 

Also, in contrast to an artificial neural network where specific network architecture --- such as the number of hidden layers and number of neurons per layer \citep[see e.g.][]{annz} --- must be chosen and potentially optimized from a very large set of possibilities, the analogous architecture of an SVM is autonomously optimized during the training process and requires only the pre-designation of a kernel function \citep{Burges98}. Further, SVM training always reaches a global solution, whereas a neural network often has multiple local minima \citep[e.g.][]{Haykin99}.  A comprehensive review of SVM algorithms is presented in e.g. \citet{Burges98} and \citet{CV95}, the latter being co-authored by the originator of the SVM concept. 

When considering SVM learning for a continuous variable such as redshift, two approaches are possible --- support vector classification (SVC) in which the output is divided into a (possibly large) number of discrete classes, or support vector regression (SVR) where the continuity of the output variable is maintained. We explored both as options with a variety of input scenarios and found that SVC performed significantly better for photo-z estimation.  

\subsection{SPIDERz package and formulation}\label{alg}

\begin{figure}[!htb]
{\includegraphics[width=8.0cm]{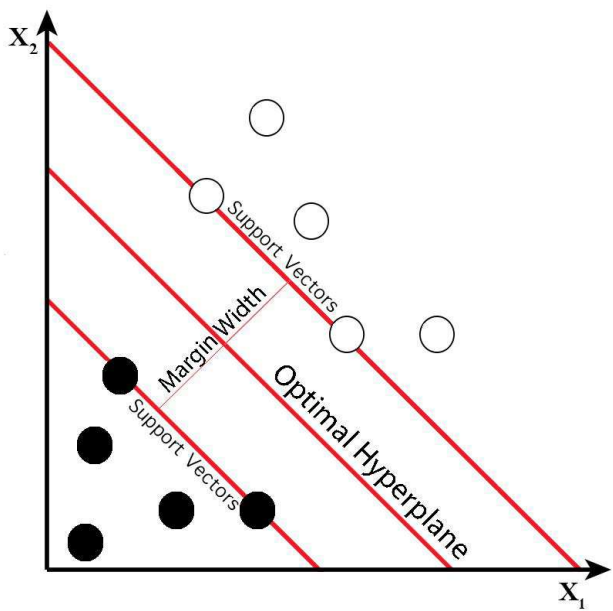}}
\caption{Depiction of linearly separable binary class system.  For visualization this figure depicts a two dimensional input space, while the input space for photo-z determination with SPIDERz is of higher dimension (the dimensions being all band magnitudes plus any other parameters).  In this depiction the two classes are filled and open circles.  The margin width is equal to $2 \over  \| \textbf{w} \|$ where $\textbf{w}$ is a vector perpendicular to the hyperplane as defined in \S \ref{alg}. }
\label{binary}
\end{figure} 

SPIDERz implements a version of SVC. We model SPIDERz on components of the general purpose LibSVM package, a library of support vector machine algorithms available in C++ and Java \citep{LIBSVM}, modified with select customizations for optimization of photo-z estimation and implementation in the IDL environment. In addition to providing discrete redshift estimates for each galaxy, SPIDERz can also output photo-z probability information (discussed below) and contains supplementary programs for preprocessing galaxy inputs and performing data analysis.

Photo-z determination with SPIDERz, and indeed any machine learning algorithm, consists of two main parts: training and evaluation. Inputs in the training process (the training set) are galaxy data consisting of band magnitudes, optional additional parameters (morphological parameters in the case discussed in this work), and known spectroscopic redshifts. The training process outputs a predictive model that can be applied to additional galaxy inputs (the evaluation set) in order to obtain photo-z estimates.

We link the galaxies in the training set to their known redshift values by dividing them into redshift bins and assigning each a representative class label.  Each galaxy and its corresponding $n$ parameters (band magnitudes and potentially morphological parameters) are represented in $n$-dimensional {\it input space} as a galaxy input vector with $n$ coordinates. 
The objective of the SVC training process is to construct a discriminative hyperplane that linearly separates galaxy vectors in the training set in such a way that maximizes the distance between galaxy vectors from opposing classes. Those vectors located closest to the separating hyperplane are termed support vectors (SVs). For data sets with classes that are not linearly separable by an unbending hyperplane in the $n$-dimensional input space, as turns out to be the case in photo-z estimation and most practical scenarios, class separation requires the mapping of input vectors $\textbf{x}$ from input space to a higher $N$-dimensional {\it feature space} with some function $\phi$ : $\textbf{x} \mapsto \phi(\textbf{x})$. Once an optimal hyperplane solution for the training set is obtained, unlabeled galaxy vectors (those in the evaluation set) are classified according to their location in feature space with respect to the separating hyperplane. 

\begin{figure}[!htb]
\resizebox{\hsize}{!}{\includegraphics{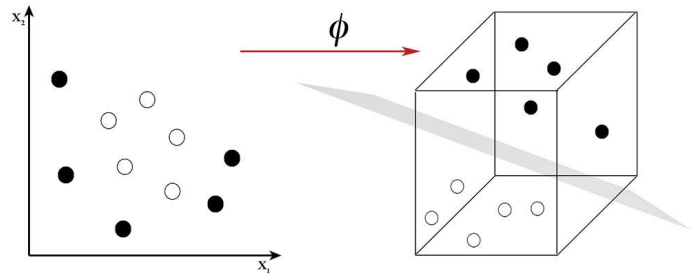}}
\caption{Conceptual visualization of binary class system which is not linearly separable in the input space (left) but is separable by a hyperplane when mapped to a higher dimensional feature space (right) with some function $\phi$ : $\textbf{x} \mapsto \phi(\textbf{x})$. }
\label{mapping}
\end{figure} 

For training a system with $m$ distinct classes ($m$ redshift bins in this case), we use a so-called ``one against one'' or ``pairwise coupling'' approach \citep{HL02,Knerr90} that divides the multi-class system into a series of $m(m-1) \over 2$ separate binary classification problems. The hyperplane optimization problem is solved separately for each pair, and the unique hyperplane solutions (also called decision functions) are consolidated and collectively output as a predictive model.  In the evaluation process, photo-z estimations are obtained by passing each galaxy vector in the evaluation set through the predictive model consisting of the $m(m-1) \over 2$ different binary classification problems solved in the training process. If a single-valued redshift prediction is desired, in the simplest implementation, which we employ in this work, the class to which a particular galaxy is most assigned becomes its final predicted class (or redshift) value (or maximum probability value --- see below). 

To explore the algorithmic procedures for determining the decision hyperplane in such a binary classification case, let us picture a linearly separable binary class system like the one depicted in Figure \ref{binary}.  For $i = 1,...,l$,  each input vector $\textbf{x}_i \in \textbf{R}^2$ has an associated class label $y_i  \in [-1,1]$, where $y_i=1$ for one class of objects and $y_i=-1$ for the other class of objects. The optimal hyperplane is definted as as $y_i(\textbf{w}\cdot\textbf{x}_i + b) = 0$, where $\textbf{w}$ is a vector perpendicular to the hyperplane and $b$ is a scalar such that the distance from the origin to the hyperplane is $|b| \over  \| \textbf{w} \|$. The following constraint is imposed on the support vectors (SVs), which are the data points located closest to the hyperplane in the space:
\begin{equation}
 y_i(\textbf{w}\cdot\textbf{x}_i + b) = 1 
\label{positiveSVs}
\end{equation}
It is important to note that the dot product $\textbf{w}\cdot\textbf{x}_i$ is a distance measure, and since $\textbf{x}_i$ and $y_i$ are strictly defined, our SV definition is effectively specifying a scale for $\textbf{w}$ and $b$. Working with these constraints, the decision function indicating the location of data points in this space can be stated as 
\begin{equation}
y_i(\textbf{w}\cdot\textbf{x}_i + b)  \geq 1, 
\label{decision_function}
\end{equation}
which we can use to define a binary classifier for this system:
\begin{equation}
\textbf{sgn} ( y_i(\textbf{w}\cdot\textbf{x}_i + b) ).
\label{decfun}
\end{equation}
Once $\textbf{w}$ is specified, this decision function is sufficient to classify objects in the linearly separable two class scheme. As previously discussed, the optimal hyperplane solution is the one in which the margin between SVs of opposing classes is maximized.  The form of the decision function sets the width of the margin to $2 \over  \| \textbf{w} \|$, therefore our objective is to minimize $\|\textbf{w} \|$, or equivalently ${1 \over 2}\|\textbf{w} \|^2$.  Finding an extremum of ${1 \over 2}\|\textbf{w} \|^2$ subject to the constraints imposed on the decision function presents a problem in the calculus of variations with a Lagrangian function
\begin{equation}
L(\alpha_i) = {1 \over 2} \|\textbf{w} \|^2 -  \sum_i {\alpha_i [y_i(\textbf{w}\cdot\textbf{x}_i + b)  - 1]}
\label{ohp1}
\end{equation}
for which the minimizing conditions are given by the Euler-Lagrange equations.  Following \citet{CV95} one finds
\begin{equation}
\textbf{w} = \sum_{i} \alpha_i y_i \textbf{x}_i
\label{weq}
\end{equation}
subject to the constraints
\begin{equation}
  \begin{cases}
\alpha_i > 0  & \quad \text{for SVs } \\
    \alpha_i = 0 & \quad \text{for non-SVs} \\
  \end{cases}
\end{equation}
\begin{equation}
\sum_{i}\alpha_i y_i = 0. 
\label{alphaconst}
\end{equation}
Thus $\textbf{w}$ can be expressed as a linear combination of SVs, and it's clear that non-SVs are irrelevant to the optimal hyperplane solution.  Finding the optimal hyperplane is now a matter of finding the corresponding $\alpha_i$ values for each SV that minimize the Lagrangian of equation \ref{ohp1}, now with the conditions in equation \ref{weq} inserted:
\begin{equation}
L(\alpha_i) = {1 \over 2} \sum_{i} \sum_{j}\alpha_i \alpha_j y_i y_j \textbf{x}_i \cdot \textbf{x}_j - \sum_{i} \alpha_i .
\label{ohp2}
\end{equation}

Let us now consider binary class systems that are not linearly separable in their original input space, like each of the $m(m-1) \over 2$ binary classification systems solved by SPIDERz.  For such systems, class separation requires mapping galaxy vectors $\textbf{x}$ from an $n$-dimensional input space to a higher $N$-dimensional feature space with some function $\phi$ : $\textbf{x} \mapsto \phi(\textbf{x})$.  A visualization of such a mapping for a simplified scenario is presented in Figure \ref{mapping}. Since the Lagrangian stated in equation \ref{ohp2} depends on dot products of input vectors living in the same input space, we represent the dot product of mapped vectors with a kernel function:
\begin{equation}
K({\textbf{x}_i,\textbf{x}_j}) = \phi(\textbf{x}_i)\cdot\phi(\textbf{x}_j).
\label{H-S-}
\end{equation}
To solve for valid kernel functions that can successfully perform input vector dot products in $N$-dimensional feature space, \citet{CV95} applied Mercer's theorem to the Hilbert-Schmidt theory  concerning dot product expansions in Hilbert space, finding that one should use symmetric and positive semi-definite kernel functions of the form:
\begin{equation}
K({\textbf{x}_i,\textbf{x}_j}) = \langle \phi(\textbf{x}_i) | \phi(\textbf{x}_j) \rangle     
\label{positive semi-definite}
\end{equation}
or equivalently,
\begin{equation}
K({\textbf{x}_i,\textbf{x}_j}) = \phi(\textbf{x}_i)^T\cdot\phi(\textbf{x}_j)      
\label{basic_kernel}
\end{equation}
where $\phi(\textbf{x}_i)^T\cdot\phi(\textbf{x}_j)$ is the inner product of the mapped vectors. There are different variants of SVM kernels, however we obtained best results with the radial basis function (RBF) kernel   
\begin{equation}
K({\textbf{x}_i,\textbf{x}_j}) =   e^{-\gamma||\textbf{x}_i-\textbf{x}_j||^2}, \gamma > 0
\label{rbf}
\end{equation}
where $\gamma$ is a scaling factor of the distance measure between input vectors $\textbf{x}_i$ and $\textbf{x}_j$ in the $N$-dimensional feature space, and whose value is assigned before training.  The RBF kernel, along with other kernels meeting conditions discovered by \citet{CV95}, allows the mapping of dot products to feature space to be calculated without explicitly determining the function $\phi$.  The free parameter $\gamma$ determines the topology of the decision surface in the feature space --- a low value sets a geometrically complicated boundary while a high value sets a geometrically simpler one with greater potential for misclassifications.

\begin{figure}[!htb]
{\includegraphics[width=8.0cm]{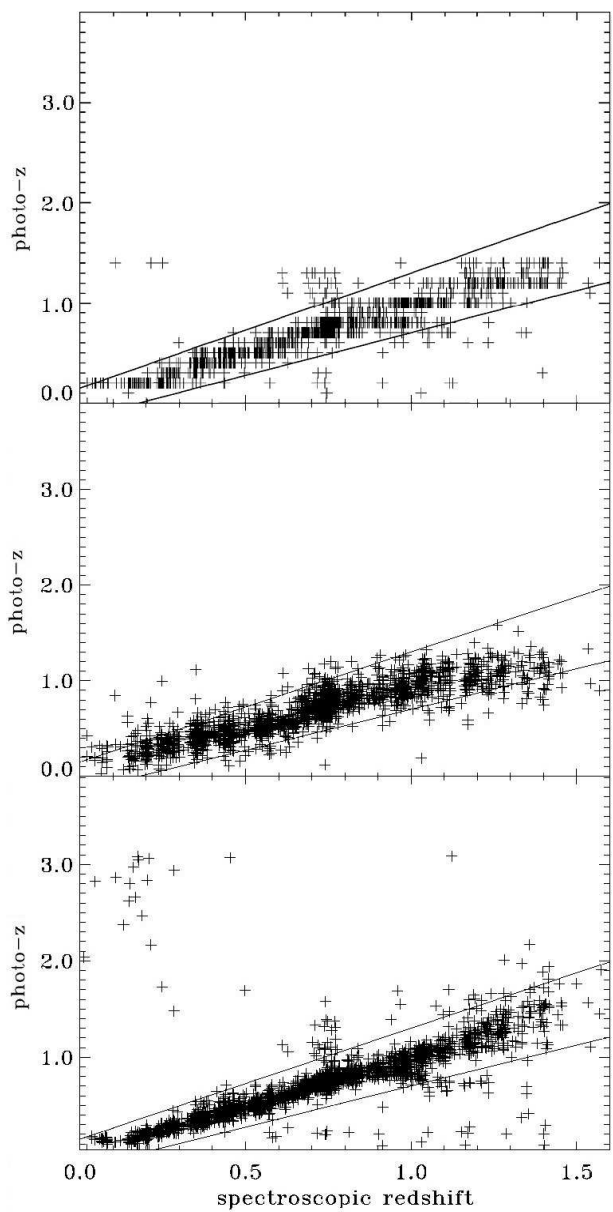}}
\caption{ {\bf TOP:} The estimated photo-z versus the actual redshift, as determined by SPIDERz, for the AEGIS data set discussed in \S \ref{AEGIS} with no morphological parameters included.  The training set is formed from 700 galaxies and the evaluation set, for which the results are plotted, consists of the remaining 1912 galaxies.  `Outliers' in a determination are defined by equation \ref{erroreq}, shown as those outside of the two diagonal lines. 
{\bf MIDDLE:} The photo-zs for the same galaxies, as estimated with the custom neural network as reported in \citet{NNP}.
{\bf BOTTOM:} The photo-zs for the same galaxies as estimated with the Lephare template fitting code \citep{lephare}.  The template fitting method has a lower scatter for non-outliers, but a larger number of outliers than the SVM and neural network for these galaxies.  However it should be noted that in contrast to empirical methods, template fitting methods such as LePhare which consider template spectra evolved at a wide range of redshifts are at a potential disadvantage when estimating photo-zs in a limited redshift range. }
\label{photovsspec}
\end{figure} 

Due to the size of the training set and number of galaxy characteristics contained in each training vector, it is not computationally practical to map the galaxy vectors to a feature space in such a way as to obtain a perfect linear separation between classes. Instead we employ a soft-margin approach \citep{CV95} in which a prospective hyperplane solution is not required to inerrantly separate class values, but rather penalized for instances of misclassification during training (mapped inputs falling on the wrong side the separating hyperplane).  This consideration is addressed with a cost function
\begin{equation}
f_C = C \, \sum_i \xi_i,
\label{costfn}
\end{equation}
where $\xi_i = 0$ if the training vector is correctly classified, and for incorrectly classified training vectors $\xi_i$ is the distance from the misclassified point to the margin boundary described by the SVs of the correct class.  The cost function is added as an additional term with additional constraints into the Lagrangian of equation \ref{ohp1}:  
\begin{equation}
L(\alpha_i,\xi_i) = {1 \over 2} \|\textbf{w} \|^2 + C \, \sum_i \xi_i  -  \sum_i {\alpha_i [y_i(\textbf{w}\cdot\textbf{x}_i + b)  - 1 + \xi_i]} 
\label{ohp3}
\end{equation}
The Euler-Lagrange equations then yield conditions that again allow the restatement of the Lagrangian as in equation \ref{ohp2} now with the additional constraint that $\alpha_i = C$ for the nonzero $\alpha_i$.  Thus as $C$ gets larger, a wider range of misclassifications are tolerated and the margin of class separation is larger, but those misclassifications that are not tolerated are penalized more.  $C$ is therefore known as a ``regularization parameter.''  

The training process now requires the minimization of the Lagrangian of equation \ref{ohp2} constrained by 
\begin{equation}
  \begin{cases}
\alpha_i = C  & \quad \text{for SVs } \\
    \alpha_i = 0 & \quad \text{for non-SVs} \\
  \end{cases}
\end{equation}
\begin{equation}
\sum_{i}\alpha_i y_i = 0. 
\label{alphaconst2}
\end{equation}
With the soft-margin approach, an optimal classifier is one that limits the number of misclassified input vectors while still maximizing the distance from the separating hyperplane, with the relative importance of each factor having a dependence on the value of $C$.

This optimization task is a so-called quadratic programming (QP) problem.  Our training process solves the QP problem using a decomposition method derived from Platt's sequential minimal optimization \citep{Platt98}, which tackles the problem by dividing it into multiple subproblems that can be analytically solved.  For this we use the algorithmic solution described in \citet{LIBSVM}.

To obtain optimized values of the free parameters $\gamma$ and $C$ that allow the classifier to construct an optimal separating hyperplane, SPIDERz conducts $v$-fold cross validation (CV) with a parameter grid search. The CV process guards against over-fitting the training set as well as approximates the performance of a predictive model on an unknown evaluation set. CV randomly separates training vectors into $v$ subsets of equal size, each of which sequentially serve as a pseudo-evaluation set for the predictive model created from a training set comprised of the remaining $v$-1 subsets.  RMS error in redshift estimation is calculated during each iteration of CV and serves as the metric of predictive power of a model obtained from training with a particular combination of $C$ and $\gamma$ values. With a grid search that performs CV using every possible combination of allowed $C$ and $\gamma$ values within a specified range and iterative step size, the program obtains optimal values of $C$ and $\gamma$ for each training set by selecting the values associated with the lowest RMS error. To expedite runtime, we first perform a course grid search to identify a rough estimate of optimal parameters, and subsequently perform a fine grid search with a smaller parameter range and step size to achieve greater precision. 

With $C$, $\gamma$, and the SVs determined by training, we can obtain a predictive model (with $m(m-1) \over 2$ realizations corresponding to each binary pair of classes, as discussed above) and use it to estimate the photo-zs of galaxies in the evaluation set. For calculating the position of an evaluation vector relative to the separating hyperplane, the kernel function is expressed in terms of support vectors  $\textbf{s}_i$
\begin{equation}
K({\textbf{s}_i,\textbf{x}}) =   e^{-\gamma||\textbf{s}_i-\textbf{x}||^2}, \gamma > 0,
\label{rbf2}
\end{equation}
and we can state a more precise formulation of the binary classifier introduced in equation \ref{decfun}:
\begin{equation}
\textbf{sgn}\bigg(\sum_{i} \alpha_i y_i K(\textbf{x}_s,\textbf{x}_i) + b\bigg).
\end{equation}

\subsection{Redshift bins, probabilities, and single-valued photo-z estimates}

Once the $m(m-1) \over 2$ unique pairwise determinations between $m$ bins of redshift have been made for a particular evaluation galaxy, the combination of these binary class estimates express what amounts to an effective probability distribution function (PDF) for each galaxy, with the relative probability of each bin proportional to the number of times the bin was chosen as the best binary solution.  This effective PDF is not continuous, but rather is resolved to the bin width level.  Alternately, if one seeks a single value photo-z prediction for a galaxy, for reasons of efficiency, simplicity, or to determine performance metrics in a similar way to previous analyses, the most probable (commonly occurring) redshift bin result could be taken as a single valued photo-z estimate for the galaxy.  For the purposes of the analyses in this work we focus on the later.

SPIDERz allows users flexibility in redshift bin size. We generally find determinations have increased accuracy and precision when smaller bin sizes are used, however the optimal bin size for any determination will be dependent on the size and nature of the training set and can be approached via trial-and-error or approximated with the bin size introduced as an additional parameter in a grid search.

The top panel of Figure \ref{photovsspec} shows the estimated photo-z versus spectroscopic redshift for the galaxies in the evaluation set of a particular determination with the AEGIS galaxy data set described in \S \ref{AEGIS} with no morphological information included.  This determination features 700 training galaxies and 1912 evaluation galaxies with a redshift bin size of 0.1.  The middle and bottom panel of Figure \ref{photovsspec} also show photo-z estimations for the same training and evaluation galaxies with a custom artificial neural network reported in \citet{NNP} and as evaluated with the Lephare template fitting method reported in \citet{lephare}.  The SPIDERz determination apparently leads to fewer catastrophic outliers than with the Lephare template fitting method or the neural network, although it shows a larger scatter among galaxies for which the photo-z estimate is close to the actual redshift.  It should be noted that template fitting methods such as LePhare, which consider template spectra evolved over a wide range of redshifts, are at a potential disadvantage when estimating photo-zs in a limited redshift range relative to empirical methods where the redshift range of the training set would match that of the evaluation set, so it is not necessarily an indication of inferior performance that the LePhare determination has a comparatively higher number of catastrophic outliers (particularly at low spectroscopic redshifts) in this case.

\section{Tests with the PHAT-1 catalog}\label{PHAT}

In order to obtain a useful comparison with other photo-z codes on publicly available data that spans a relatively large redshift range, we first focus on data available from the PHoto-z Accuracy Testing (PHAT) program, which was implemented several years ago to test and compare photo-z codes \citep{Hildebrandt10}.

\subsection{PHAT-1 dataset}\label{PHATdata}

At the time of the PHAT program, three spectroscopic and photometric catalogs were assembled, with one (designated PHAT-1) based on real galaxy data.  PHAT-1 featured a catalog of nearly 2000 galaxies with spectroscopic redshifts and 18-band photometry from the Great Observatories Origins Deep Survey (GOODS) North field.  In the catalogs which were made available, three-quarters of the spectroscopic redshifts, chosen at random, were ``blinded'' to the users with the intention that the users would utilize the other one-quarter as the training set and predict the redshifts of the blinded galaxies, with results evaluated by the PHAT authors.  As the PHAT program is no longer operational, we did not have the option of evaluating SPIDERz on the blinded galaxies.  Therefore, we use only the unblinded galaxies, and to maintain the same proportions of training and evaluation galaxies as in the PHAT program, we use one-quarter (chosen at random) as the training set and the remaining three-quarters as the evaluation set.  For a more straightforward evaluation we also ignore galaxies with missing band magnitudes, which corresponds to the ``cleaned sample'' analysis reported in \citet{Hildebrandt10} and results in a set of 374 galaxies with known redshifts ranging from $0.08 < z < 3.6$, and thus a training set with 94 galaxies and an evaluation set with 280 galaxies.

\subsection{Results with PHAT-1 data}\label{PHATresults}

\begin{figure}[!htb]
{\includegraphics[width=8.0cm]{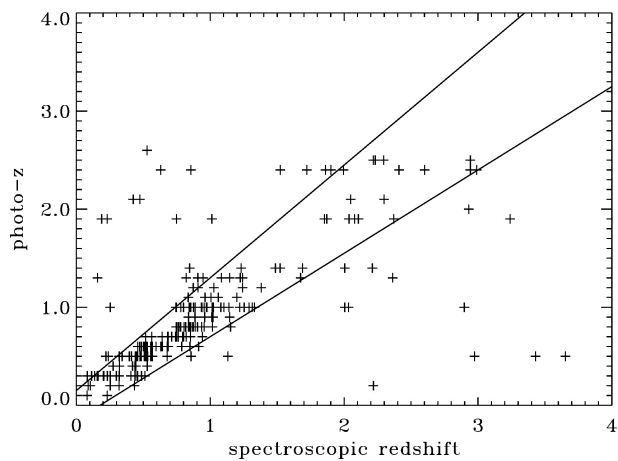}}
\caption{ The estimated photo-z as determined by SPIDERz versus the actual redshift for the PHAT-1 data set discussed in \S \ref{PHATdata} for the case of 14 photometric bands.  This determination is with a training set consisting of 94 galaxies chosen at random and an evaluation set consisting of the other 280 galaxies.  Outliers are defined by equation \ref{erroreq}, shown as those outside of the two diagonal lines.  This determination results in 15.3\% outliers and a reduced RMS of 0.067, which, despite the smaller training set available, is competitive with the results obtained by other empirical methods as shown in Table 1 and discussed in \S \ref{PHATresults}.  }
\label{PHATfig}
\end{figure}

Even with this greatly reduced training set size our results on the PHAT-1 data are comparable to many other photo-z codes.  In Table 1 we show the results with SPIDERz along with those from other empirical codes as reported in Table 5 of \citet{Hildebrandt10}.  The other empirical codes reported there are the artificial neural network package ANNz \citep{annz}, an empirical $\chi^2$ fitting code \citep{Wolf09}, a polynomial fit \citep{LY08}, and regression trees \citep{Carliles10}.  Results for the eight template fitting codes reported in that work vary from a high of 27.6\% outliers and 0.061 reduced RMS to a low of 4.7 \% outliers and 0.038 reduced RMS. In Table 1 we also show results as reported for the two best-performing template fitting codes, LePhare \citep{lephare} and BPz \citep{bpz}.  Figure \ref{PHATfig} shows the estimated photo-z versus actual redshift results of one particular determination with SPIDERz on the PHAT-1 dataset.  We note that although the results from a full 18-band analysis are reported in \citet{Hildebrandt10}, for a comparison analysis with SPIDERz, we used only 17 bands in our determinations because of a large number of missing values in one band.

\begin{table}
\caption{Results for the ``cleaned sample'' of the PHAT-1 catalog with one-quarter of galaxies used for training and the remaining three-quarters used for evaluation, shown for SPIDERz as determined in this work along with  other empirical codes and the two best performing template fitting codes as reported in \citet{Hildebrandt10}. Results for SPIDERz are shown for bin sizes of both 0.1 and 0.2 in redshift.  Outliers are defined by equation \ref{erroreq} and the reduced RMS is given by equation \ref{RMSeq} with outliers excluded.  As discussed in \S \ref{PHATdata} a much smaller training set was available for the SPIDERz determinations than was available for the others.  }
\label{tab}
\begin{tabular}{lrrrr}
\hline
  & 14-band &  & 18-band &  \\         
Code  & Outlier \% & R-RMS & Outlier \% & R-RMS \\
\hline
SPIDERz (0.1) & 15.3 & 0.067 & 18.9  & 0.059 \\
SPIDERz (0.2) & 15.3 & 0.064 & 13.9 & 0.062 \\
ANNz & 36.5 & 0.078 & 29.0 & 0.074 \\
$\chi^2$ Emp. & 13.3 & 0.066 & 15.3 & 0.067 \\
Poly Fit & 9.4 & 0.051 & 14.5 & 0.052  \\
Regress. Trees & 21.6 & 0.067 & 19.0 & 0.066 \\
\hline
BPz & 7.8 & 0.041 & 7.5  & 0.060 \\
LePhare & 4.7 & 0.038 & 4.9  & 0.040 \\
\hline
\end{tabular}
\end{table}

\section{Tests with the COSMOS bright catalog}\label{COSMOS}

In order to obtain a data set with a large number of objects, and to investigate the effects of the inclusion of morphological information in photo-z determination, we turn to the COSMOS field \citep[e.g.][]{Scoville07}, where multi-band photometry, spectroscopic redshifts, and morphological information are available for a large number of galaxies. 

\subsection{COSMOS bright  data}\label{Cosdat}

In particular, we combine photometry from the COSMOS2015 photometric catalog \citep{Laigle15}, the publicly available portion of the zCOSMOS catalog of spectroscopic redshifts \citep[e.g][]{Lilly09}, and the morphological parameters provided by \citet{Cassata}.  Using a 0.5'' matching criterion for objects, we obtain a catalog of 14365 galaxies with photometry in $u$, $B$, $V$, $r$, $i$, $z+$, $Y$, $H$, $J$, and $Ks$ bands, seven morphological parameters, and spectroscopic redshifts rated as very secure.  By construction all redshifts lie between 0 and 1.4.

Although the COSMOS2015 catalog provides photometry in a large number of optical, infrared, and UV bands, we choose to restrict our analyses to the bands listed above for several reasons.  One reason is that outside of those bands the rate of missing photometry is much higher. Another is that with data sets approaching 30 bands of photometry, the distinction between photo-z estimation and spectroscopic redshift determination is somewhat muddled, and in any case this does not represent a realistic photometric situation for upcoming large surveys such as LSST, even for subsets which would have infrared survey overlap. 

The seven morphological parameters present in the catalog are

1. The Petrosian Radius $R_P$: The mean radial distance from the center of a galaxy at which the local intensity of the light equals some multiple $\eta$ of the average intensity of light within the radius: $I(R_P) = \eta \left( { { \int_0^{R_P} I(r) 2 \pi r dr } \over {\pi {R_P}^2 } } \right)$

2. The Half Light Radius $r_{e}$: The radial distance from the center of a galaxy within which half the total light is contained.

3. The Concentration $c$: $c=5 \, log \, {{r_{80}} \over {r_{20}}}$.
This parameter defines the central density of the light distribution with radii $r_{80}$ and $r_{20}$ correspondingly 80\% and 20\% of the total light.

4. The Asymmetry $A$:  $A = { {\Sigma_{x,y} \vert I_{(x,y)}-I_{180(x,y)} \vert} \over {2\Sigma_{x,y} \vert I_{x,y} \vert} } -B_{180}. $
This parameter characterizes the rotational symmetry of the galaxy's light, with $I_{(x,y)}$ being the intensity at point (x,y) and $I_{180(x,y)}$ being the intensity at the point rotated 180 degrees about the center from (x,y), with $B_{180}$ being the average asymmetry of the background calculated in the same way.  It is the difference between object images rotated by 180$^{\circ}$.

5. The Gini coefficient $G$: \\
$G={ {1} \over {\bar{X} \, n(n-1)} } \Sigma^n_i (2i-n-1) \, X_i$,
describes the uniformity of the light distribution, with $G=0$ corresponding to the uniform distribution and $G=1$ to the case when all flux is concentrated in to one pixel.  $G$ is calculated by ordering all pixels by increasing flux $X_i$. $\bar X$ is a mean flux and $n$ is the total number of pixels.

6. $M_{20}$: $M_{20}=log \Sigma M_{i}/M_{tot}$,
is the ratio of the second order moment of the brightest 20\% of the galaxy to the total second moment.  This parameter is sensitive to the presence of bright off-center clumps.

7.  The Axial Ratio $\varepsilon$:  $\varepsilon = 1 \, - \,  {{b} \over {a}}$.
The values $a$ and $b$ are the semi-major axis and semi-minor axis of the galaxy.

These parameters are discussed at greater length in e.g. \citet{Scar06}.

\subsection{Results with COSMOS bright data}\label{Cosres}

\begin{figure}[!htb]
{\includegraphics[width=8.0cm]{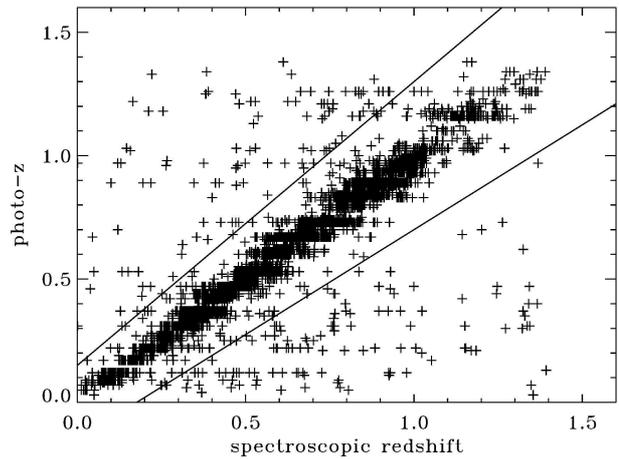}}
\caption{ The estimated photo-z as determined by SPIDERz versus the actual redshift for the COSMOS data set discussed in \S \ref{COSMOS}.  This determination was performed with ten-band photometry and a designated bin size of 0.01. The training and evaluation sets are comprised of 3000 and 11365 randomly chosen galaxies, respectively.  With this small bin size, a relatively large ratio of training set objects is needed in order to adequately bins with training objects.  Outliers in a determination are defined by equation \ref{erroreq}, shown as those outside of the two diagonal lines.  The density of points within the lines is quite high --- only 2.0\% of points lie outside of the lines as outliers.  This determination features a reduced RMS error of 0.022. }
\label{COSMOSfig2}
\end{figure} 

\begin{figure}[!htb]
{\includegraphics[width=8.0cm]{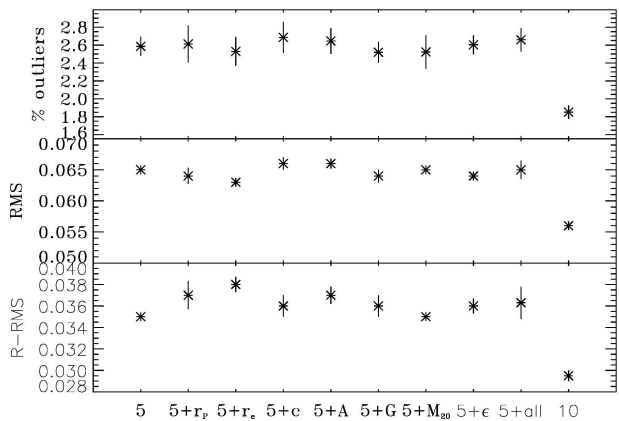}}
\caption{Percentage of outliers (TOP) along with non-reduced (MIDDLE) and reduced (BOTTOM) RMS error $\sigma_{\Delta z/(1+z)}$ in the photo-z estimation with the inclusion of different bands and morphological parameters in the COSMOS field data set discussed in \S \ref{COSMOS} as determined with SPIDERz.  The uncertainties represent the standard deviation of the values obtained from different realizations featuring different random compositions of the training and evaluation sets, as discussed in \S \ref{Cosres}. Results are shown for the five-band case, the cases of five photometric bands plus each individual morphological parameter, the case of five photometric bands plus all morphological parameters, and the ten photometric band case.}
\label{COSMOSfig}
\end{figure} 

We present several results with the COSMOS bright catalog.  First we present results just utilizing the five optical bands $u$, $B$, $r$, $i$, and $z+$, which could resemble the default situation for obtaining photometric redshifts from a very large optical survey.  We also present results using these five optical bands {\it plus} morphological parameters in order to explore whether the inclusion of morphological information can improve photo-z estimation in future large optical surveys. For completeness and comparison purposes we also present results utilizing all of the ten photometric bands listed above.  Figure \ref{COSMOSfig2} shows a plot of estimated photo-z versus actual redshift for a particular determination in the ten-band case where we have used a small bin size of 0.01 as a demonstration.  

In order to quantify the variance for different determinations, we complete six realizations of training and evaluation for every case, each with a randomized training set of 1000 galaxies and an evaluation set of the remaining 13365 galaxies, and record the number of outliers and RMS errors for the evaluation. Because the membership of the training and evaluation sets is randomized, we obtain slightly varying numbers of outliers and RMS errors with each evaluation.  Figure \ref{COSMOSfig} shows the averaged number of outliers and RMS errors for determinations performed using five bands, five bands plus the inclusion of individual shape parameters, five bands plus the inclusion of all shape parameters, and lastly ten bands.  We see that the inclusion of shape information does not result in a statistically significant improvement in the fraction of outliers, RMS, or reduced RMS as compared to the five-band only case.  We discuss this further in \S \ref{disc}.  As expected, estimates with the ten-band case are significantly better than with the five-band case.  We also note that enlarging the training set size beyond 1000 galaxies for this data did not yield an appreciable improvement in either the fraction of outliers or the RMS errors when a binsize of 0.1 was used.

It is potentially interesting to compare the accuracy obtained with SPIDERz to that reported by the COSMOS collaboration's own photo-z analysis on similar data.  With the former, in the ten-band case, we achieve an average of 1.8\% outliers and an average reduced RMS of 0.022.  The numbers reported by the latter which most closely correspond to the former are from the ``bright spectroscopic redshifts'' sample reported by the COSMOS collaboration in \citet{Laigle15} where they achieve 0.5\% outliers and a reduced RMS of 0.007 with a method that is based on the LePhare \citep{lephare} template fitting code.  However, it must be noted that the results in \citet{Laigle15} are obtained using up to 30 bands of photometry from ultraviolet to infrared including some quite narrow (< 300 \AA) bands.  The two results then are difficult to compare directly, as the analysis with SPIDERz is disadvantaged by having fewer bands.

\section{Tests with the COSMOS and 3D-HST overlap data}\label{COS-HST}

The COSMOS bright data set discussed in \S \ref{COSMOS} has the advantage of containing many thousands of galaxies, but the disadvantage of being limited to relatively low redshifts. To obtain a data set of real galaxies with publicly available spectroscopic redshifts that contains higher redshift sources we use spectroscopic redshifts from the 3D-HST survey performed with the Hubble Space Telescope and reported in \citet{Momcheva16} that overlap with COSMOS photometry and morphology.  This results in a data set of 3048 galaxies, of which 206 (6.8\%) have $z > 2$ and 537 (17.6\%) have $ z > 1.5$.  These galaxies have photometry and morphological information as discussed in \S \ref{Cosdat}.  We note that this is the largest redshift range for which the inclusion of morphological information for photo-z determination has been tested, and that potentially morphological information may manifest an advantage in data sets with larger redshift ranges that is not present in narrower ranges.

\begin{figure}[!htb]
{\includegraphics[width=8.0cm]{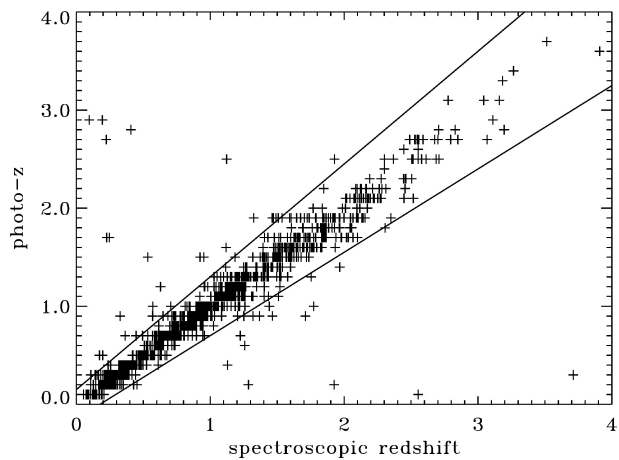}}
\caption{ The estimated photo-z as determined by SPIDERz versus the actual redshift for the COSMOSx3D-HST data set discussed in \S \ref{COS-HST}.  This determination is with the ten-band photometric data and a training set consisting of 1200 galaxies chosen at random and an evaluation set consisting of the other 1848 galaxies, with a bin size of 0.1.  Outliers in a determination are defined by equation \ref{erroreq}, shown as those outside of the two diagonal lines.  The density of points within the lines is quite high --- only 2.6\% of points lie outside of the lines as outliers.  This determination features a reduced RMS error of 0.04. }
\label{COSHSTfig2}
\end{figure} 

\begin{figure}[!htb]
{\includegraphics[width=8.0cm]{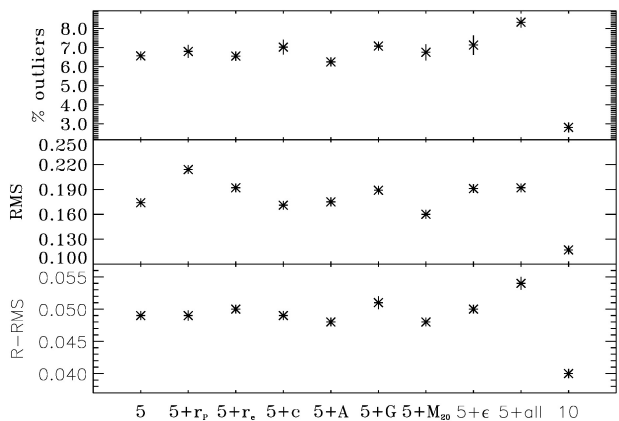}}
\caption{Percentage of outliers (TOP) along with non-reduced (MIDDLE) and reduced (BOTTOM) RMS error $\sigma_{\Delta z/(1+z)}$ in the photo-z estimation with the inclusion of different bands and morphological parameters in the COSMOSx3D-HST data set discussed in \S \ref{COS-HST} as determined with SPIDERz.  The uncertainties represent the standard deviation of the values obtained from different realizations featuring different random compositions of the training and evaluation sets, as discussed in \S \ref{Cosres}. Results are shown for the five photometric band case, the cases of five photometric bands plus each individual morphological parameter, the case of five photometric bands plus all morphological parameters, and the ten photometric band case.}
\label{COSHSTfig}
\end{figure} 

As with the COSMOS bright data in \S \ref{Cosres}, for the COSMOSx3D-HST data we present several results.  First we present results just utilizing the five optical bands $u$, $B$, $r$, $i$, and $z+$, which could resemble the default situation for obtaining photometric redshifts from a very large optical survey.  We also present results using these five optical bands {\it plus} morphological parameters in order to explore whether the inclusion of morphological information can improve photo-z estimation in future large optical surveys.  For completeness and comparison purposes we also present results utilizing all of the ten photometric bands listed above.  Similarly, we complete six realizations of training and evaluation for every case, each with a randomized training set of 1200 of the galaxies and an evaluation set of the remaining 1848 galaxies, and record the number of outliers and the RMS errors for the redshift determination for the evaluation set.  For this data we increase the proportion of training set galaxies in an effort to train the model with a greater representation of the relatively scarce highest redshift population.  Again, each randomized realization produces a slightly different number of outliers and RMS errors.  Figure \ref{COSHSTfig2} shows a plot of estimated photo-z versus actual redshift for one particular determination in the ten-band case. Figure \ref{COSHSTfig} shows the number of outliers and RMS errors for data sets composed of five bands, five bands plus the inclusion of individual shape parameters, five bands plus the inclusion of all shape parameters, and ten bands.  We see again that the inclusion of shape information does not result in a statistically significant improvement in the fraction of outliers, RMS, or reduced RMS as compared to the five-band only case.  As expected, estimates with the ten-band case are significantly better than with the five-band case.  

We note that SPIDERz performs remarkably well on this data set, including on the high redshift objects, considering that a small fraction of the galaxies are high redshift.  

\section{Tests with AEGIS data} \label{AEGIS}

\subsection{AEGIS data}\label{Adata}

For comparison with previous results obtained with a neural network algorithm \citep{NNP} we also perform tests on observations of the Extended Groth Strip from the the All-wavelength Extended Groth Strip International Survey (AEGIS) data set \citep{EGS}, which contains photometric band magnitudes in $u$, $g$, $r$, $i$, and $z$ bands from the Canada-France-Hawaii Telescope Legacy Survey \citep[CFHTLS,][]{Gwyn08}, imaging from the Advanced Camera for Surveys on the Hubble Space Telescope \citep[HST/ACS,][]{K07}, and spectroscopic redshifts from the DEEP 2 survey using the DEIMOS spectrograph on the Keck telescope.  The limiting $i$ band AB magnitude of the CFHTLS survey is 26.5, while that of HST/ACS is 28.75 in $V$ (F606W) band, and that of DEEP2 is 24.1 in $R$ band.  

A total of 2612 galaxies spanning redshifts from 0.01 to 1.57, with a mean redshift of 0.702 and a median of 0.725, and $i$ band magnitudes ranging from 24.43 to 17.62, are in the data set used here.  The redshift distribution of this particular set of galaxies arises because of the intentional construction of the portion of DEEP2 spectroscopic catalog within the AEGIS survey to have roughly equal numbers of galaxies below and above $z$=0.7; therefore it is not an optimized training set for a generic photometric data evaluation set, although a more optimized training set for any given photometric data evaluation set could be constructed from it.  

For shape information we use the following morphological parameters, previously derived in other works from the HST/ACS imaging data in two bands, V (F606W) and I (F814W):\\
1. The Half Light Radius $r_e$, \\
2. The Concentration $c$,\\
3. The Asymmetry $A$,\\
4. The Gini coefficiet $G$,\\
5. $M_{20}$, and\\
6. The Axial Ratio $\varepsilon$,\\ 
all as defined in \S \ref{Cosdat}, as well as

7. The Smoothness $S$:  
$S = { {\Sigma_{x,y} \vert I_{(x,y)}-I_{S(x,y)} \vert} \over {2\Sigma_{x,y} \vert I_{x,y} \vert} } -B_{S} $.
The smoothness is used to quantify the presence of small-scale structure in the galaxy.  It is calculated by smoothing the image with a boxcar of a given width and then subtracting that from the original image.  In this case $I_{(x,y)}$ is the intensity at point (x,y) and $I_{S(x,y)}$ is the smoothed intensity at (x,y), while $B_S$ is the average smoothness of the background, calculated in the same way.  The residual is a measure of the clumpiness due to features such as compact star clusters.  In practice, the smoothing scale length is chosen to be a fraction of the Petrosian radius.

and

8. The S\'ersic power law index $n_s$ where the S\'ersic profile has a form $\Sigma(r) = \Sigma_e\,e^{-k \vert (r/r_e)^{(1/n_s)}-1 \vert}$ \citep[e.g.][]{GD05}, where $\Sigma_e$ is the surface brightness at radius $r_e$ and $k$ is defined such that half of the total flux is contained within $r_e$.

Parameters $c$, $A$, $S$, $G$, and $M_{20}$ are determined in \citet{Lotz08}, while $\varepsilon$, $n_s$, and $r_e$ are determined by \citet{Griffith11} using the Galfit package \citep{Peng02,Hausler07}.

For comparison with a previous result in this case we form principal components of the morphological parameters for this data set.  Principal components \citep[e.g.][]{Jolliffe02} are the result of a coordinate rotation in a multi-dimensional space of possibly correlated data parameters into vectors with maximum orthogonal significance.  The first principal component is along the direction of maximum variation in the data space, the second is along the direction of remaining maximum variation orthogonal to the first, the third is along the direction of remaining maximum variation orthogonal to both of the first two, and so on.  The morphological principal components are given as linear combinations of the eight morphological parameters in Table 1 of \citet{NNP}.  

A further discussion of the particular mapping from morphological to principal components is available in \S 3 of \citet{NNP}.  As mentioned there the first principal component is well correlated with galaxy type, and correlations persist through several of the other principal components.  These correlations hinted that the morphology may provide an additional handle on the photo-z estimation, since outliers often occur because a spectral feature (such as a break) of one galaxy type at a given redshift may be seen by the observer to be at the same wavelengths as a spectral feature of another galaxy type at different redshift.  Thus it was an intriguing hypothesis that morphological information indicative of galaxy type may help break this degeneracy.  

\subsection{Results with AEGIS data} \label{simlumf}

As in \S \ref{Cosres} and \S \ref{COS-HST} we complete six realizations of training and evaluation for every case, for this data with a randomized training set of 700 of the galaxies discussed in \S \ref{Adata} and an evaluation set of the remaining 1912 galaxies, and record the number of outliers and the RMS error for the redshift determination for the evaluation set.   Here as well because in each realization the membership of the training and evaluation sets varies, each realization for a given input parameter set produces a slightly different number of galaxies in the evaluation set that are outliers and a slightly different RMS error.  For comparison, the template fitting results reported by \citet{lephare} using band photometry only give 5\% outliers and a (non-reduced) RMS error of $\sigma_{\Delta z/(1+z)} =.1881$ for this sample.  This error is dominated by the catastrophic outliers (Figure \ref{photovsspec}), and clearly visually drops to below that of SPIDERz if outliers are excluded.

\begin{figure}
{\includegraphics[width=8.0cm]{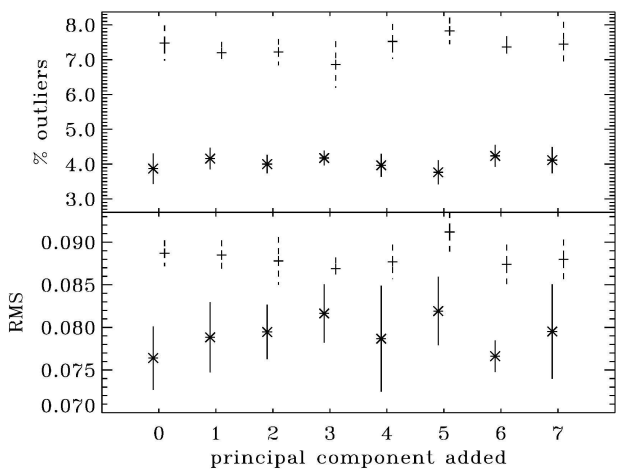}}
\caption{Percentage of outliers (TOP) and non-reduced RMS error $\sigma_{\Delta z/(1+z)}$ (BOTTOM) in the photo-z estimation with the inclusion of different individual principal components of the morphological parameters with the AEGIS data set discussed in \S \ref{AEGIS}, shown for SPIDERz (solid lines and star points) and for the custom neural network described in \citet{NNP} (dashed lines and cross points).  The uncertainties represent the standard deviation of the values obtained from different realizations featuring different compositions of the training and evaluation sets, as discussed in \S \ref{simlumf}.  }
\label{indpcs}
\end{figure} 

\begin{figure}
{\includegraphics[width=8.0cm]{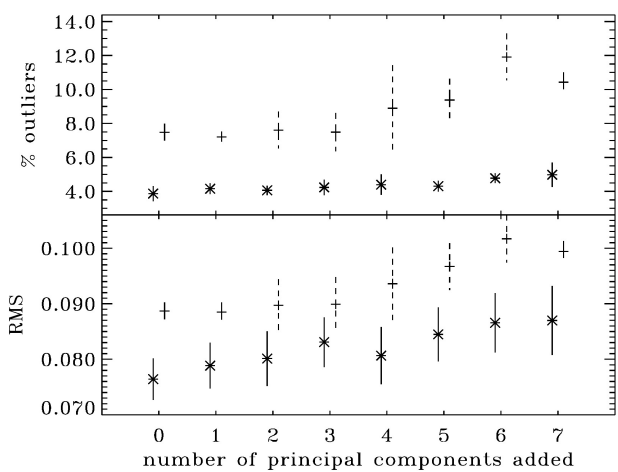}}
\caption{Percentage of outliers (TOP) and non-reduced RMS error $\sigma_{\Delta z/(1+z)}$ (BOTTOM) in the photo-z estimation  with the inclusion of multiple principal components of the morphological parameters with the AEGIS data set discussed in \S \ref{AEGIS}, starting with none, adding the first morphological principal component, then the first and the second principal components, and so on.  Results are shown for SPIDERz (solid lines and star points) and for the custom neural network described in \citet{NNP} (dashed lines and cross points).  The uncertainties represent the standard deviation of the values obtained from different realizations featuring different compositions of the training and evaluation sets, as discussed in \S \ref{simlumf}.}
\label{multpcs}
\end{figure} 

Figure \ref{indpcs} shows the number of outliers and RMS error for the inclusion of the seven different principal components individually.  Figure \ref{multpcs} shows the number of outliers and RMS error for the inclusion of multiple principal components, starting with none, then adding in the first, then adding in the first and second, then adding in the first through third, and so on.  We do not show the reduced RMS in this case because it is not available for comparison in the neural network determination.  In each figure, the error bars correspond to the standard deviation of the number of outliers or RMS scatter in the different realizations.  We note that the last principal component (PC8) should by definition contain minimal significant variation in the morphological parameters, so we do not include it in the analysis.

It is apparent that the SVM algorithm of SPIDERz results in a significant decrease in the number of outliers and RMS error compared to the artificial neural network algorithm previously tested. In addition, the standard deviation for different realizations in the number of outliers is generally smaller with the SVM method, allowing us to determine that adding morphological information in the form of individual principal components of the morphological parameters does not provide a statistically significant decrease in the average number of outliers or the RMS error, while adding multiple principal components increases the number of outliers and the RMS error.

\section{Discussion}\label{disc}

We have developed a custom SVM classification algorithm for photometric redshift estimation in the IDL environment.  The  package, SPIDERz, is available to the community.  It outputs for each galaxy both an effective distribution of probabilities (in bins of redshift) for each galaxy's photo-z and a single-valued, most likely predicted photo-z bin, with the bin size chosen by the user.  In this analysis for practicality of evaluating metrics such as outliers and RMS errors we use only the single-valued photo-z prediction.

We compare results obtained with SPIDERz to those previously reported for other codes with the PHAT-1 catalog in \S \ref{PHATresults}.  We see that SPIDERz performs comparable to other empirical methods on this catalog, noting that as discussed in \S \ref{PHATdata} far fewer training galaxies were available for SPIDERz.  On this latter point, having such a reduced number of training galaxies available is a significant disadvantage for SPIDERz or any empirical method in a dataset such as PHAT-1 where proportionally few of the galaxies are higher redshift.  In the context of SVC as discussed in \S \ref{alg}, as the training data in certain redshift bins becomes sparse, there may be insufficient support vectors available for those bins in the binary classifications for effective hyperplane solutions.

SPIDERz can naturally include additional parameters beyond band magnitudes.  We note that an SVM is in a sense an unbiased way of determining the relative strength of the correlations of a set of input parameters with the output parameter, and that this SVM algorithm is designed to treat all input parameters on an equal footing, thus providing for a convenient method for investigating the inclusion of additional inputs beyond photometry. 

In order to both explore a much larger dataset and examine the effects of the inclusion of morphological parameters on photo-z estimation, we form a data set of 14365 galaxies with photometry, morphological parameters, and spectroscopic redshifts from the COSMOS field as discussed in \S \ref{COSMOS}.  We find that while SPIDERz performs relatively well on this data, that the inclusion of morphological information in the form of morphological parameters does not improve the number of outliers or RMS errors in photo-z estimation over the case of five  optical photometric bands only.  As expected, using ten photometric bands reduces the number of outliers and RMS errors considerably relative to five bands.

To study the performance of SPIDERz and the effects of including morphological parameters over a larger redshift range we form a data set of 3048 galaxies from the overlap of COSMOS photometry and morphology with 3D-HST spectra as discussed in \S \ref{COS-HST}.  We find SPIDERz performs quite well on this data, including on higher redshift galaxies, even though these galaxies form a small fraction of the  training set.  We again find that the inclusion of morphological information in the form of morphological parameters does not improve the number of outliers or RMS errors in photo-z estimation over the case of five optical photometric bands only, and again using ten photometric bands reduces the number of outliers and RMS errors considerably relative to five bands.

For comparison with a previous determination with a neural network algorithm, we also estimate photo-z for a data set which consists of 2612 galaxies with five optical band magnitudes, reliable spectroscopic redshifts, and principal components of eight morphological parameters, discussed in \S \ref{AEGIS}.  We again find that the inclusion of morphological information does not significantly decrease the number of outliers or RMS error in photo-z estimation.  Previously \citet{NNP} found compatible results with the neural network estimation method.  We do note that the inclusion of multiple principal components resulted in a diminished performance in photo-z estimation with the neural network algorithm due to the addition of noise while this had less of an effect on the performance of SPIDERz, indicating that perhaps SVMs are less susceptible to the negative effects of adding noise in this situation.

The particular findings regarding the inclusion of shape information thus evidence some robustness in regard to different data sets, consideration of morphological parameters or principal components, and consideration of an SVM or neural network method.  We conclude, therefore, that these results are likely applicable to all empirical methods with this redshift range and photometry restricted to the visible and near infrared bands.  Any gain that may arise in an empirical photo-z determination from correlations between morphology and redshift is overwhelmed by the additional noise introduced.  It is likely that any correlations between the morphological parameters and the galaxy type are degenerate to some extent with the correlations between galaxy type and galaxy colors.  The possibility remains that under a less input-blind and more complicated scenario where, for example, the regularization parameter for misclassification (see equation \ref{costfn}) varies with the redshift, the additional noise contained in morphological information could be made less consequential.  A potentially more straightforward possibility is that morphological information along with probability distribution considerations could be used only to flag galaxies more likely to be catastrophic outliers, which we will explore in a future work.

\begin{acknowledgements}
The authors acknowledge data supplied previously by B. Gerke, M. Shmakova, R.L. Griffith, and J. Lotz.  Based in part on observations made with the NASA/ESA Hubble Space Telescope, obtained from the Data Archive at the Space Telescope Science Institute, which is operated by the Association of Universities for Research in Astronomy, Inc., under NASA contract NAS 5-26555.  Funding for the DEEP2 survey has been provided by NSF grants AST-0071048, AST-0071198, AST-0507428, and AST-0507483. Some of The data presented herein were obtained at the W. M. Keck Observatory, which is operated as a scientific partnership among the California Institute of Technology, the University of California and the National Aeronautics and Space Administration. The Observatory was made possible by the generous financial support of the W. M. Keck Foundation. The DEEP2 team and Keck Observatory acknowledge the very significant cultural role and reverence that the summit of Mauna Kea has always had within the indigenous Hawaiian community and appreciate the opportunity to conduct observations from this mountain. 
\end{acknowledgements}

\end{document}